\input phyzzx
\newcount\mongocount
\mongocount=1
\def\Figure#1#2#3{
% \boxit{
      \vbox to #3in{\hsize=#2in
        \vfil
%\special{ps::[begin]
%          save 10 dict begin /Figure exch def
%          currentpoint translate
%          /showpage {} def
%        }
%        \special{ps: plotfile #1}
         \includegraphics{#1}
%        \special{ps::[end]
%        clear Figure end restore
%        }
    }
% }
}
\def\figcap#1#2{
\vtop{\tenpoint\singlespace
\hsize=#1in\smallskip\noindent Figure\ \ \the\mongocount.\ \  #2
\global\advance\mongocount by 1\bigskip}}
\def\mongofigure#1#2#3#4#5{\centerline{\Figure{#1}{#2}{#3}
\figcap{#4}{#5}}}

\hoffset=0.375in
\overfullrule=0pt

\def\au{{\rm AU}}

\def\dol{{D_{\rm OL}}}
\def\dls{{D_{\rm LS}}}
\def\dos{{D_{\rm OS}}}

\def\ms{{\rm Machos}\ }

\def\kpc{{\rm kpc}}

\def\absz{|\zeta|}
\def\kms{{\rm km}\,{\rm s}^{-1}}
\def\ms{{\rm m}\,{\rm s}^{-1}}

\def\epsil{{\epsilon}}
\def\epsilt{{\epsilon_{thres}}}
\twelvepoint
\font\bigfont=cmr17
\centerline{\bigfont Detection Rates for Close Binaries Via Microlensing}
\bigskip
\centerline{{\bf B. Scott Gaudi}}
\smallskip
\centerline{{\bf Andrew Gould}\footnote{1}{Alfred P.\ Sloan Foundation Fellow}}
\smallskip
\centerline{Dept of Astronomy, Ohio State University, Columbus, OH 43210}
\smallskip
\centerline{e-mail gaudi@payne.mps.ohio-state.edu}
\smallskip
\centerline{e-mail gould@payne.mps.ohio-state.edu}
\bigskip
\centerline{\bf Abstract}

Microlensing is one of the most promising methods of reconstructing
the stellar mass function down to masses even below the hydrogen-burning limit.
The fundamental limit to this technique is the presence of unresolved binaries,
which can in principle significantly alter the inferred mass function.  Here we quantify
the fraction of binaries that can be detected using microlensing, considering
specifically the mass ratio and separation of the binary.  We find that
almost all binary systems with separations
greater than $b \sim 0.4$ of their combined Einstein ring radius are detectable assuming
a detection threshold of $3\%$.
For two M dwarfs, this corresponds to a limiting separation of $\gsim 1 \au$.  
Since very few observed M dwarfs have companions at separations $\lsim 1 \au$, we
conclude that close binaries will probably not corrupt the measurements of the mass
function.  We find that
the detectability depends only weakly on the mass ratio.   For those events for 
which individual masses can be determined, we 
find that binaries can be detected down to $b \sim 0.2$.  
 
\singlespace 
%\doublespace

\bigskip
Subject Headings: 
binaries -- gravitational lensing 
\smallskip
\centerline{submitted to {\it The Astrophysical Journal}: June 17, 1996}
\centerline{Preprint: OSU-TA-16/96}  
\endpage
\normalspace
\chapter{Introduction}

Four surveys are currently discovering microlensing events 
towards the Large Magellenic Cloud and the galactic bulge (Alcock et al.\ 1996; Aubourg
et al.\ 1995; Udalski et al.\ 1994; Alard 1996).  While the
initial goal of these surveys was to determine the fraction of the halo that
is composed of massive compact halo objects, the possible returns on these
surveys are much broader.  In particular, it may soon be possible to measure
the mass function of the lenses.  

Traditional methods of measuring the stellar mass function are restricted to luminous
objects.  Thus these methods can only be applied to stars above the hydrogen-burning limit,
and are restricted to sparse samples near this limit.
Microlensing overcomes
this limitation because the effect is due to the mass of the lens, not its intrinsic luminosity.
Thus microlensing samples can extend mass function measurements 
beyond the hydrogen-burning limit.

In general it is not possible to measure the masses of individual microlenses.  This is
because the only parameter that yields any information about the lens is the timescale $t_e$,
given by,
$$
t_e = { {r_e}\over{v}},\eqn\timescale
$$
where $v$ is the transverse velocity of the lens relative to the observer-source
line of sight, and $r_e$ is the Einstein
ring radius,
$$
r_e^2 = {{4GM}\over{c^2}}\dos\, z(1-z), \,\,\, z={{\dol}\over{\dos}}.
\eqn\rade
$$
Here $\dol$, $\dos$, and $\dls$ are the distance between the observer, lens, and source, 
and $M$ is the mass of the lens.  Thus $t_e$ 
is a complicated function of the quantities of interest: the mass, velocity, and distance
of the lens.  There are
two basic methods of acquiring additional information.  The first is using parallax 
to measure the projected Einstein radius of the lens, 
${\tilde r}_e = (\dos/\dls) r_e$, either by considering the parallax caused by
the motion of the Earth (Gould 1992; Alcock et al.\ 1995; Buchalter \& Kamionkowski 1996), or
by employing a parallax satellite (Refsdal 1966; 
Gould 1995a; Boutreux \& Gould 1996; Gaudi \& Gould 1996).  The second method is using proper
motion information to measure the angular Einstein radius, $\theta_e = 4GM/c^2{\tilde r}_e$ 
(Gould 1994; Nemiroff \& Wickramasinghe 1994; Witt \& Mao 1994; Gould \& Welch 1996).
Combining these two
pieces of information yields the mass, distance and velocity of the lens (Gould 1996a and
references therein).  Gould (1995b) estimates that one can expect $\sim 100$ giant events
towards the galactic bulge per year.  A parallax satellite would be able to measure
parallaxes for $\sim 70\%$ of these events (Gaudi \& Gould 1996), and 
$\sim 15\%$ of events could yield proper motions with current technology (Gould 1996a).  Thus
one might expect to obtain full information for $\sim 15$ events per year.

The fundamental limitation to using microlensing to reconstruct the mass function comes
from the issue of unresolved binaries.  If one assumes that the individual masses 
measured are due to single lenses without considering unresolved binaries, 
the reconstructed mass function will be biased toward
large masses.  This issue has been studied for the stellar mass function as determined
from counts of luminous stars in the solar
neighborhood (Reid 1991; Kroupa et al.\ 1991), and it has been shown that unresolved binaries
can significantly alter the inferred mass function, in particular leading to an
underestimation of the numbers of low-mass stars.  Unresolved binaries could pose a similar problem
for microlensing.  It is thus important to quantify the detectability of binaries 
from microlensing.

Binary events can be divided into
three basic classes according to the separation, $b$, in units of the
Einstein ring radius: wide binaries ($b \gg 1$), 
intermediate binaries ($b \sim 1$), and
close binaries ($b \ll 1$).  Although it may be difficult to determine 
the frequency of wide binaries from microlensing experiments, 
these objects pose no problem for reconstructing
the mass function because the light curve for each member is unaffected by the presence of 
the other.  
Similarly, intermediate binaries pose no difficulty because they give rise to events that
deviate dramatically from those of single lenses, and hence are easily distinguished. 
However, close binaries are problematic in that they can masquerade as point lenses.
In this way microlensing differs from traditional methods of detecting binaries:  
the closer a companion is to a luminous star, the larger the induced orbital motion
and hence the {\it{easier}} it is to detect spectroscopically.

The goal of this paper is to quantify the fraction of binary microlensing events for
which the binarity of the lens is detectable.  We specifically focus on close binaries, 
for which $b \le 1$.  In
particular we quantify the smallest separation that can be resolved for a majority
of events given specified observing parameters.       
We also discuss the effects of the mass ratio of the binary components and
effects of finite source size on the detectability.

\chapter{Observed Frequencies of Close Binaries}

Before calculating detection rates, we first review what is known about the
frequency of close binaries, specifically for low-mass stars and brown dwarfs, where
microlensing is most useful.  Unfortunately, there is no information about the
binarity of objects with masses below the hydrogen-burning limit, the regime of greatest
interest. The most relevant observed sample is of the stars just above
the hydrogen-burning limit.  Various surveys of local, late-type dwarf stars have
been made with the aim of discovering unseen companions.  Precise radial
velocity measurements are the most sensitive to low-mass, close binaries.
Marcy \& Benitz (1989) obtained radial velocity measurements of M dwarfs in the
solar neighborhood, with precisions of $\sim 200\, \ms$, allowing
detection of companions with masses 
$\gsim 0.01 M_{\odot}$.  Fischer \& Marcy (1992) examined this
sample, and found that out of 62 primaries, only three have
a companions with separations of $\lsim 1 \au$.  They estimate a detection
probability of $86\%$ for this range.  For two M dwarfs, $1 \au$ corresponds to a 
separation in units of the Einstein ring radius of $b \sim 0.4$.  
Thus approximately $3/(.86 \times 62) 
\sim 6\%$ of M dwarfs in the solar neighborhood have companions with $b < 0.4$.  
Although the solar neighborhood may not be a perfectly representative sample, it appears that 
binaries with separations $b < 0.4$ are not common.  As we show below,
with observations of reasonable photometric precision,  
almost all binaries with separations $b > 0.4$ are detectable.
Moreover, we show that for events where the mass can actually be measured, binarity is
detectable even at substantially smaller separations.              
Thus close binaries are unlikely to be a major source of error for reconstructing the mass
function via microlensing.

\chapter{Binary Lensing Formalism}

Consider a binary lens system.
The Einstein ring radius for the binary
system is given by equation \rade, where $M$ now denotes the 
total mass of the binary.
We will normalize all subsequent lengths and masses to $r_e$ and $M$. 
Using complex coordinates, we denote the position of the source with 
respect to the center of mass of the binary as $\zeta = \xi + i\eta$, and
the position of the component masses $m_1$ and $m_2$ as
$z_1$ and $z_2$.  The image positions, $z=x+iy$, are then given by 
(Witt 1990),
$$
\zeta = z + {{m_1\over{\bar{z_1}}-{\bar{z}}}} +{{m_2\over{\bar{z_2}}-{\bar{z}}}}.
\eqn\lenseq
$$ 
The magnification, $A_i$, of each image is given by the Jacobian
of the transformation \lenseq, evaluated at the image position,  
$$
A_i = \left.{1\over{ |{\rm{det}} J|}}\right|_{z=z_i},\, \, \,
{\rm{det}}J = 1 - {{\partial\zeta}\over{\partial{\bar z}}}
\overline{{{\partial\zeta}\over{\partial{\bar z}}}}.
\eqn\mag
$$
If the images are unresolved, the total magnification is given by the sum of the
individual magnifications, $A = \sum A_i$.
The set of source positions where the magnification
is formally infinite, given by the condition ${\rm{det}} J = 0$, 
define closed curves called caustics.  Five images are
created if the source is inside a caustic, three if the source is outside.
It is those regions near the caustics where the magnification from the
binary deviates most dramatically from that of a point lens.  

In practice, equation \lenseq\ is solved numerically in order to determine the
image positions, and these positions, together with equation \mag, 
are then used to calculate the total magnification.

\chapter{Detection Rates}

After normalizing to $r_e$ and $M$, there are two parameters 
that determine the lensing structure of a binary: $b$, the binary separation, and the 
mass ratio, $q=m_1/m_2$.   To analyze how the magnification of a 
binary lens deviates from that of a point mass lens for a specified $q$ and $b$, 
we define $\epsil$, the 
excess magnification over a single lens,
$$
\epsil = {{A-A_0}\over {A_0}},
\eqn\excess
$$
where $A_0$ is the magnification of a point lens with mass equal to the
total mass of the binary, $M$, and located at the center of mass of the binary,
$$
A_0 = {{\absz^2 +2}\over{\absz\, (\absz^2 +4)^{1/2}}}.
\eqn\anaught
$$

\FIG\one{
Contours of excess magnification, $\epsil =0, \pm 0.03,\pm 0.10$,
for four values of $b$, the binary separation in units of the Einstein ring,
and $q$, the ratio of the masses of the binary components. Positive contours are bold. 
}
\FIG\two{
Contours of detection probability as a function of binary separation and
mass ratio for $\epsilt = 0.10$.  Contours have equal spacings of $10\%$.  A binary is
considered detected if $|\epsil| > 0.10$ at any point during the event. 
}
\FIG\twob{
Caustic curves, defined as the locus of points
in the source plane where the magnification is formally
infinite, for binary separations of $b=1.0, 2^{-1/2}, 0.6$.  
}
\FIG\three{
Contours of detection probability as a function of binary separation and
mass ratio for $\epsilt = 0.03$.  Contours have equal spacings of $10\%$.  A binary is
considered detected if $|\epsil| > 0.03$ at any point during the event.      
}
\FIG\four{
Contours of excess magnification, $\epsil =\pm 0.03,\pm 0.10$,
for a source of radius $\rho = 0.03$ for events with $|\zeta| \le 0.06$ for
binary separations of $b=0.25,0.20,0.15,0.10$.  
Positive contours are bold.  
}

\FIG\five{
Contours of detection probability as a function of binary separation 
for $\epsilt = 0.03 {\rm{(solid)}}, 0.10 {\rm{(dashed)}}$.  
The source size is $\rho = 0.03$.  A binary is
considered detected if $|\epsil| > \epsilt$ at any point during the event. 
}

We then calculate $\epsil$ as a function of the source position, 
$(\xi ,\eta)$, and draw contours of $\epsil = \pm 3\%, \pm 10\%$.  
These excess magnification contours define regions where
the magnification of the binary lens deviates from that of the single lens
by $3\%$ and $10\%$. Figure \one\ shows contours of $\epsil$ for 
$q= 0.1, 0.4, 0.7, 1.0$ and $b=0.2, 0.3, 0.4, 1.0$. 

\topinsert
\mongofigure{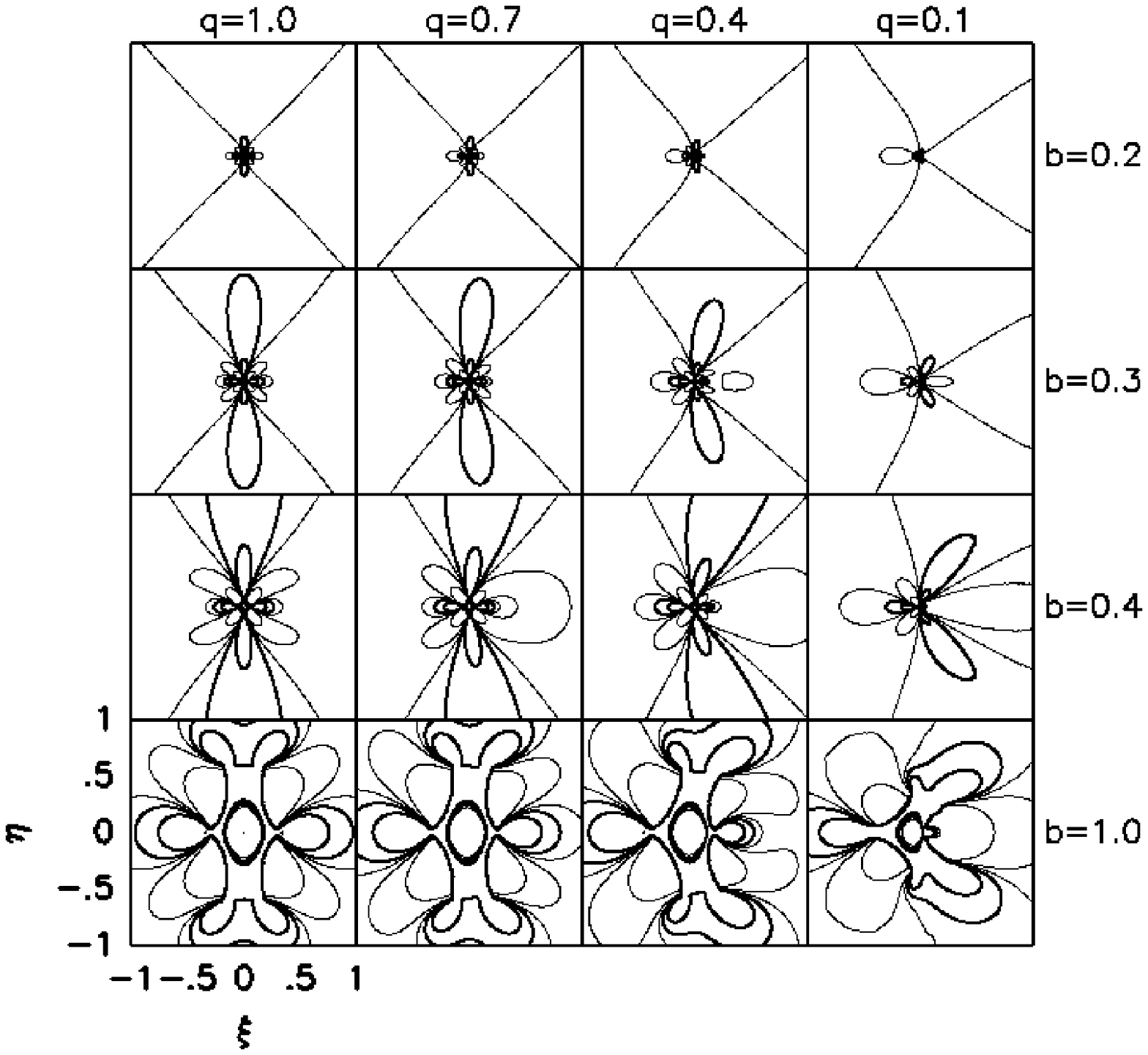}{6.4}{6.5}{7.0}
{
Contours of excess magnification, $\epsil =0, \pm 0.03,\pm 0.10$,
for four values of $b$, the binary separation in units of the Einstein ring,
and $q$, the ratio of the masses of the binary components. Positive contours are bold.
}
\endinsert

Ignoring higher-order effects (unresolved light from unmagnified sources, 
finite source, etc.), there
are six parameters that characterize a binary lens event.  Two of these are
intrinsic to the lens: $b$ and $q$.  Three are purely geometrical factors
that describe the lens trajectory: $t_0$, the time of closest approach of the source
and the center of mass of the lens,
$\beta$, the separation between the source and center of mass at $t_0$,
 and $\theta$, the angle that the lens trajectory makes with the 
projected binary axis.  The final parameter is the timescale of the event,
given by equation \timescale.    
A lensing event will be a straight line through
the maps of $\epsil$, specified by these six parameters.  We want to know,
given any event being observed with a specified sampling rate, $n_{meas}$,  
what is the probability of distinguishing a binary with parameters
$q$ and $b$ from a point-mass lens.      
This means we must consider all possible $\beta$ and $\theta$.  Since
$t_0$ has no effect on the detection probability, we ignore it.  For each
possible trajectory, we ask whether, at each measurement, 
$|\epsil| \ge \epsilt$, where $\epsilt$ is the given detection threshold.  
If this requirement is met, we consider that the binary has been detected.
The distributions of $\beta$ and $\theta$ are flat.   We therefore
integrate over all trajectories with 
$0 \le \beta \le 1$ and $0 \le \theta \le 2\pi$. The probability for 
detecting the binary is simply the ratio of the number of
events for which the binary was detected to the total number of
trial events.  
Comparing the panels of Figure \one, it is apparent that the binary detection 
probability depends much more strongly on $b$ than $q$.     
For $b=1.0$, nearly all possible event trajectories will cross contours 
of $3\%$, and therefore the fraction of events for which the binary is 
detected is $\sim 1$, whereas for $b=0.2$, the fraction of events for which
the binary is detected is $\ll 1$.  For $b=0.4$, a significant fraction of
trajectories will still cross contours of $3\%$.  This implies that the 
binary detection probability must decline rapidly from $b=0.4$ to $b=0.2$. 

\topinsert
\mongofigure{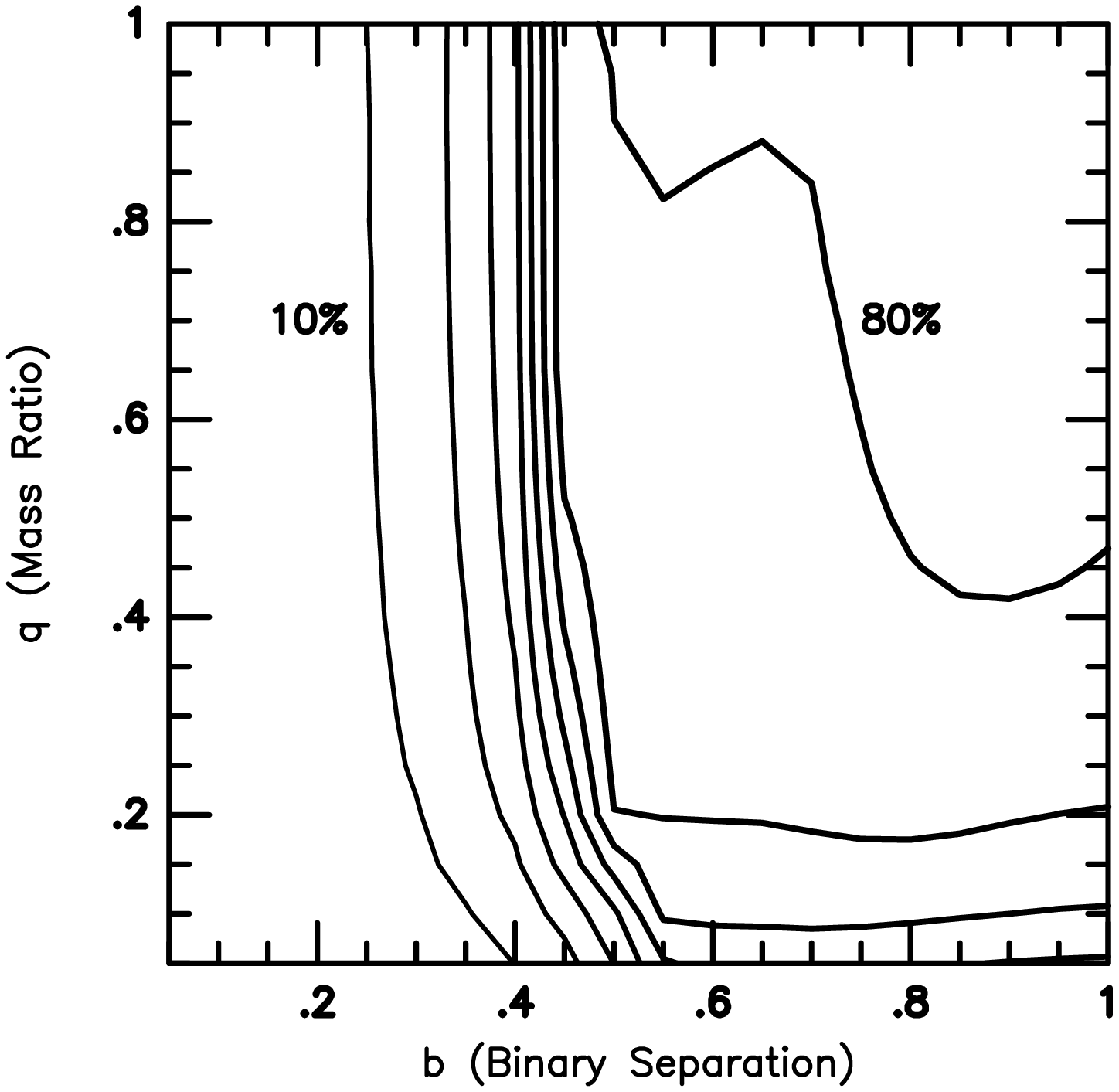}{6.4}{6.0}{6.4}
{
Contours of detection probability as a function of binary separation and
mass ratio for $\epsilt = 0.10$.  Contours have equal spacings of $10\%$.  A binary is
considered detected if $|\epsil| > 0.10$ at any point during the event.
}
\endinsert
\topinsert
\mongofigure{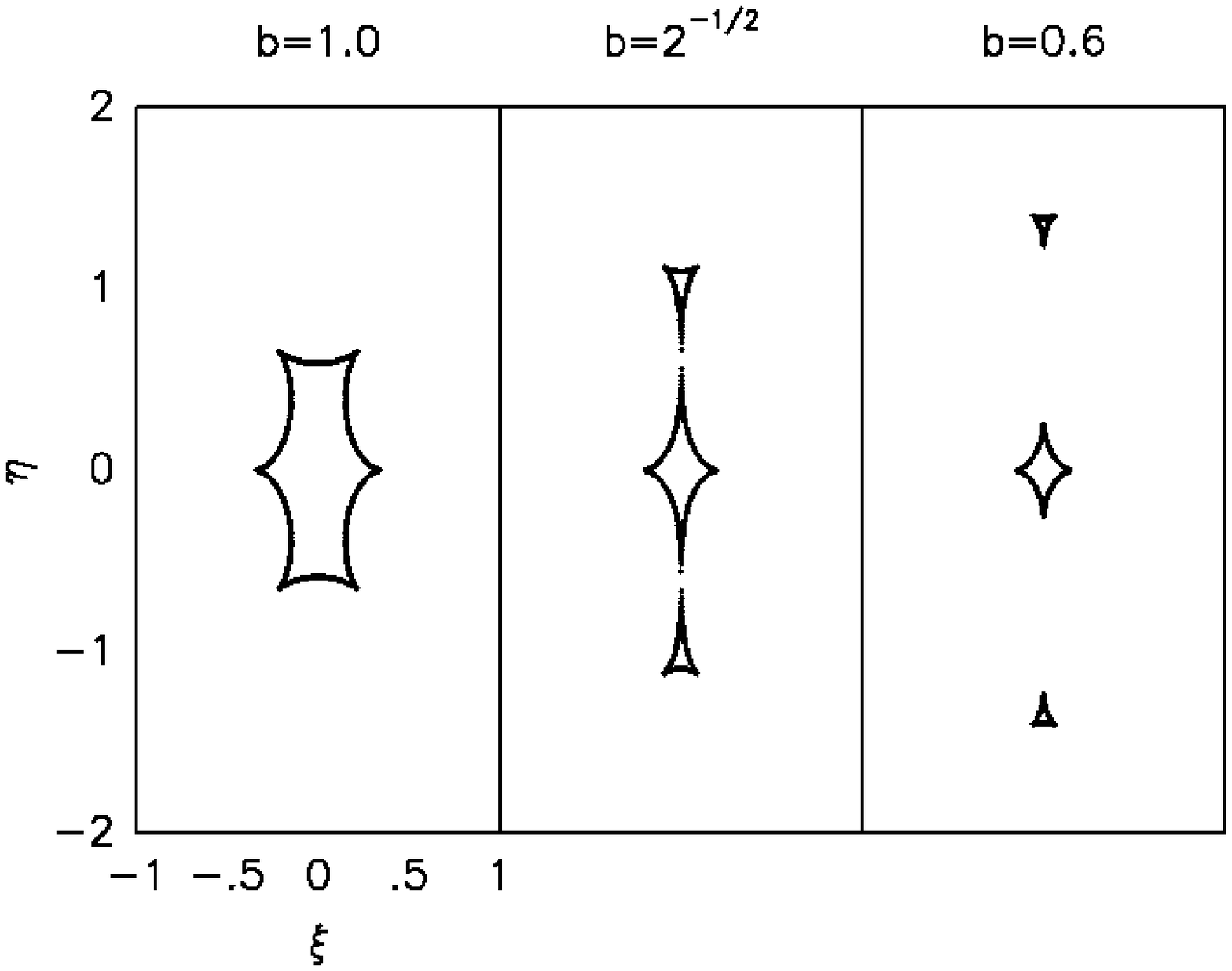}{6.4}{6.0}{7.0}
{
Caustic curves, defined as the locus of points
in the source plane where the magnification is formally
infinite, for binary separations of $b=1.0, 2^{-1/2}, 0.6$.  
}
\endinsert
\topinsert
\mongofigure{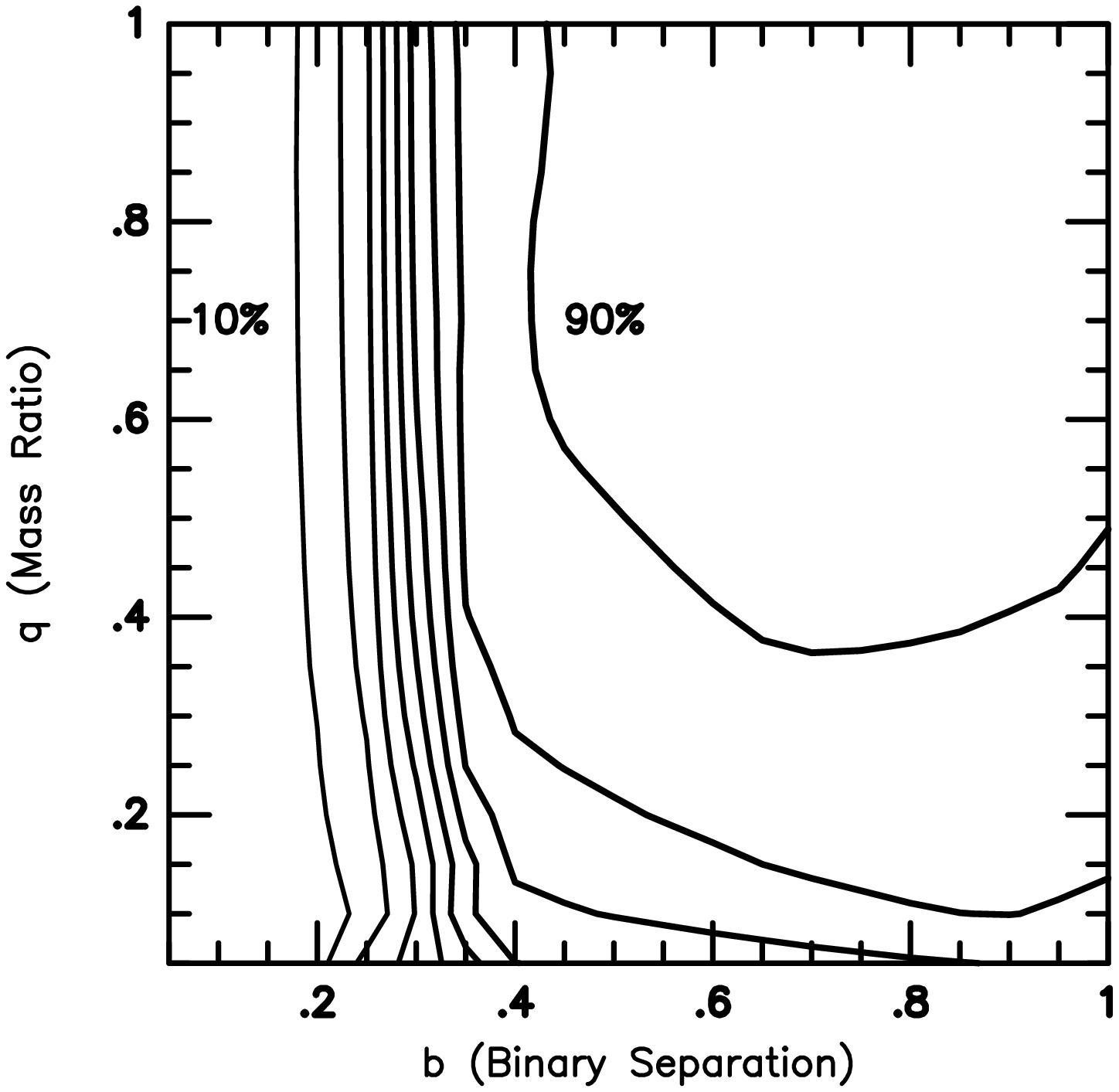}{6.4}{6.0}{6.4}
{
Contours of detection probability as a function of binary separation and
mass ratio for $\epsilt = 0.03$.  Contours have equal spacings of $10\%$.  A binary is
considered detected if $|\epsil| > 0.03$ at any point during the event.      
}
\endinsert

In order to calculate binary detection probabilities, we adopt parameter values
$\epsilt = 3\%$ and $10\%$, and $n_{meas} = 25$ measurements per $r_e$.
For $\dos \sim 8 \, \kpc$ and $\dol \sim 4 \, \kpc$,
$$
r_e \sim 4 \, {\rm{AU}}\left({{M}\over{M_{\odot}}}\right)^{1/2}.
\eqn\rescale
$$
Thus for a binary consisting of two $\rm{M}$ stars, $M \sim 0.4 M_{\odot}$,
$r_e \sim 3 \,{\rm{AU}}$.    
Assuming $v \sim 200\, \kms$, then $t_e \sim 25\, {\rm{days}}$.  Therefore
$n_{meas}$ corresponds to one measurement per day.  In fact, our results
are rather insensitive to choice of $n_{meas}$, since, from Figure \one,
for $b \ge 0.3$, the regions of $\epsil > 0.1$ are all 
larger than $\sim 0.1 r_e$, and therefore the binary will be detected for
any $n_{meas} > 10$.  Furthermore, current follow-up surveys have
temporal resolutions much greater than those used in our calculations.

Figure {\two} shows contours of binary detection probability as a function of
$b$ and $q$ for $\epsilt = 10\%$. As expected, $P$ is much more sensitive to 
$b$ than $q$, and declines rapidly for $b \lsim 0.5$. 
Excess magnification contours of $\epsil = 0.10$
follow closely the structure of the caustics.  Thus we can examine the
structure of the caustics to understand Figure \two.  In Figure {\twob}, we show the
caustics for three binary separations.  For separations of $0.7 \lsim b \le 1.0$, 
there is only one caustic, and the binary detection probability is roughly given 
by the cross section of this caustic integrated over all angles $\theta$.  At
a binary separation of $b=2^{-1/2} \simeq 0.7$, the caustics splits into three parts. 
  The main, diamond-shaped caustic is 
located near the center of mass, and the two smaller triangle-shaped caustics are
on the $\eta$ axis symmetrically above and below the main caustic.  As $b$ decreases
from $b=0.7$, the secondary caustics become smaller and move farther from the center of mass,
but the $\epsil = 0.1$ contour still extends between the primary and secondary caustics.  
At $b \sim 0.5$, the $\epsil= 0.1$ contour breaks, and no longer extends between the
primary and secondary caustics.  Thus the binary detection probability shown in
Figure {\two} exhibits a
sharp break at $b \sim 0.5$, and declines rapidly for $b < 0.5$, as the primary caustic
shrinks.

Figure {\three} shows contours of binary detection probability as a function of
$b$ and $q$ for $\epsilt = 3\%$. Again, $P$ is much more sensitive to 
$b$ than $q$, and declines rapidly for $b < 0.4$.  The reasons for the structure of
Figure {\three} are similar to those for $\epsilt = 3\%$, but note from 
Figure {\one} that the $\epsil = 0.03$ 
contours follow the structure of the caustics less closely, and thus the break
in the detection probability occurs at a smaller value of $b$.

\chapter{Orbital Motions}

To understand the effect of the orbital motion of the binary on the detection
probability, we define a parameter $\psi$, which describes the amount the
binary rotates during the event,
$$
\psi = 2\pi {{t_e}\over P},
\eqn\psieq
$$
where $P$ is the period of the binary.  Using Kepler's laws, and assuming
face-on circular orbits and $\dos \sim 8\, \kpc$, $\psi$ can be written as,
$$
\psi = { \left( {{15\,\kms}\over{v}} \right)}{ \left( {{M}\over{M_{\odot}}} \right)}^{1/4}
{ \left[ {4z (1-z) } \right]}^{1/4} b^{-3/2}.
\eqn\psieval
$$
For $v \sim 200\, \kms$, $z=1/2$, and $M \sim 0.4 M_{\odot}$, this 
becomes $\psi \sim 0.06 b^{-3/2}$.  
Therefore, for $b > 0.3$, the binary will rotate by $\psi \lsim 20^{\circ}$.  
Thus the rotation of the binary during an event is
small for most events, and will not change significantly the binary 
detection probabilities.  This result is borne out quantitatively in numerical simulations
that we have performed, but which we do not report in detail.

\chapter{Finite Source Effects}

In order to utilize microlensing as a method to reconstruct a mass function, 
one must be able to gather additional information for each individual event.
One of the two necessary pieces of information is the proper motion of the lens,
$\mu = v/\dol$.  For events with relatively small Einstein rings, (which are
typically associated with the low-mass lenses considered here), $\mu$ can
be measured primarily when the source passes very close to the lens.
The light curve then deviates from that of a point source, and this deviation can then be used
to determine $\mu$.

It is therefore interesting to restrict consideration to those events for 
which finite source effects
must be taken into account, 
and to ask what is the probability of detecting a binary if these effects are present.
Since the majority of source stars for these events will be giants, we consider a source 
of average giant radius $R=22R_{\odot}$ (Gould 1995b).  For $\dos =8\,\kpc$, 
$\dol =4\,\kpc$, and $M \sim 0.4 M_{\odot}$, this corresponds to a 
projected distance on the source plane, 
normalized to $r_e$, of $\rho = 0.03$.  Gould \& Welch (1996) 
estimate that, using optical/infrared photometry,
proper motions could be measured
when $\beta \lsim 2\rho$.  We therefore restrict our
attention to those trajectories for which $\beta \le 0.06$.  In general, when a 
source crosses a caustic, the magnification will deviate dramatically from
that of a point mass lens, and the binary will be easily detectable.  
However, if the size of the caustic, which we will denote as $w$, 
is smaller than that of the source,
$$
w \lsim \rho,
\eqn\caus
$$
then finite source effects will mask the binary magnification signature.  
If the caustics is
very much smaller than the source, $w \ll \rho$, 
then the light curves for the binary and point-mass lenses
will appear nearly identical when finite source effects are included.   
For $b \le 0.3$, the primary caustic is diamond-shaped, and it can be shown
analytically that,
 $$
w \simeq b^2/2.
\eqn\caussize
$$  
Combining equations {\caus} and {\caussize}, we find that finite source
effects will partially mask the binary for $b \lsim 0.25$.

\topinsert
\mongofigure{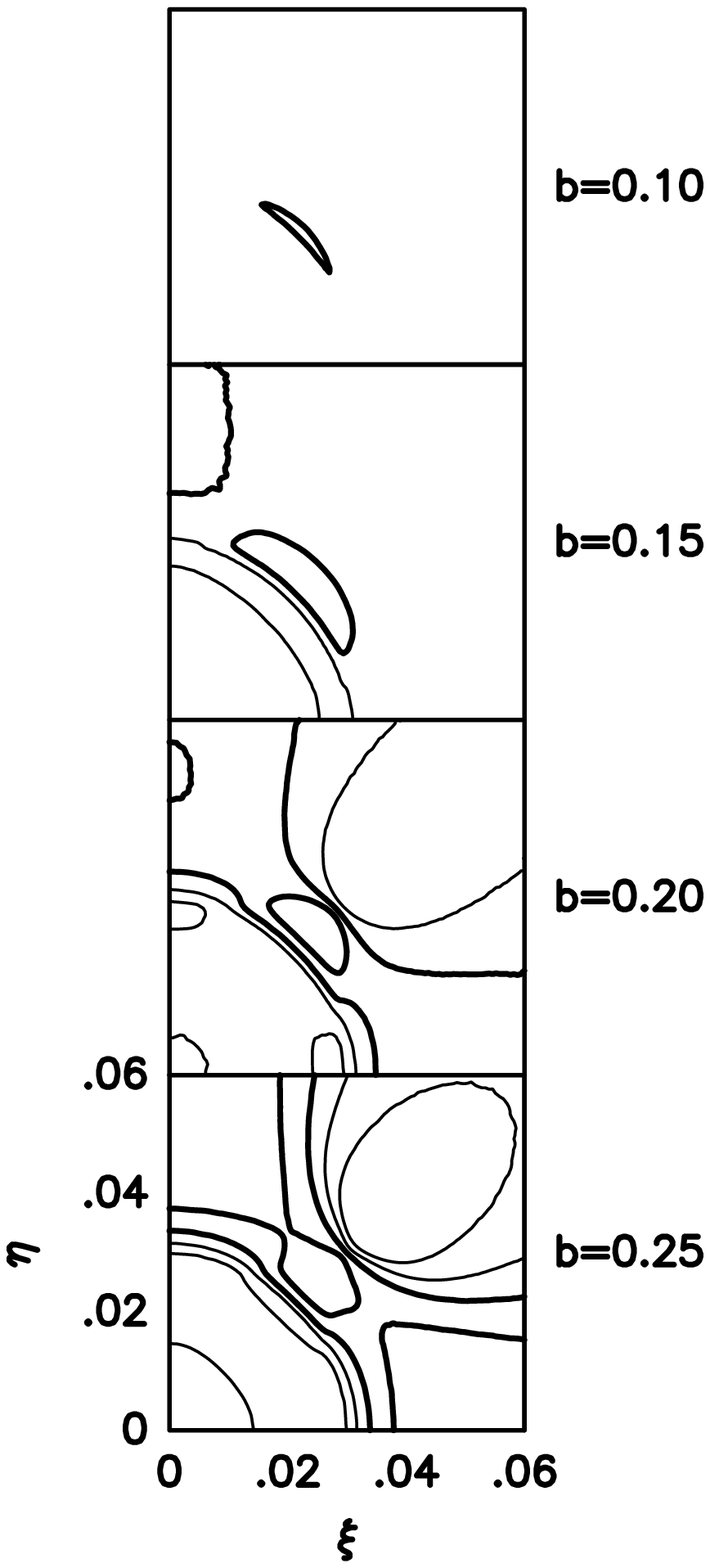}{6.4}{8.0}{6.4}
{
Contours of excess magnification, $\epsil =\pm 0.03,\pm 0.10$,
for a source of radius $\rho = 0.03$ for events with $|\zeta| \le 0.06$ for
binary separations of $b=0.25,0.20,0.15,0.10$.  
Positive contours are bold.  
}
\endinsert
\topinsert
\mongofigure{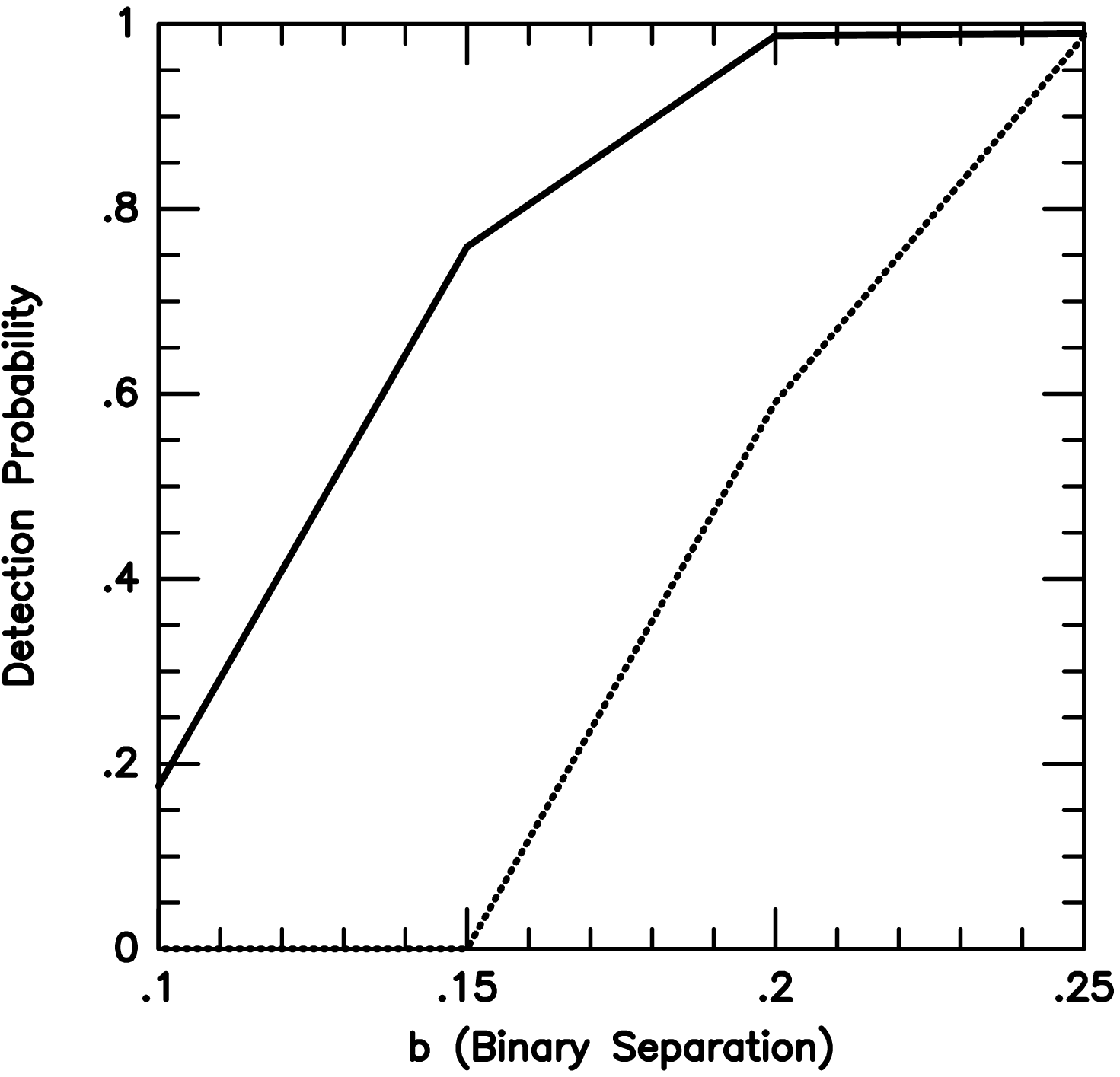}{6.4}{6.0}{6.4}
{
Contours of detection probability as a function of binary separation 
for $\epsilt = 0.03 {\rm{(solid)}}, 0.10 {\rm{(dashed)}}$.  
The source size is $\rho = 0.03$.  A binary is
considered detected if $|\epsil| > \epsilt$ at any point during the event. 
}
\endinsert

To be more quantitative, we again calculate the excess magnification, $\epsil$,
over a point lens [c.f.\ eq.\ {\excess}], where now $A$ and $A_0$ are the 
magnifications of a finite source of radius $\rho = 0.03$.  In order to calculate
this magnification, we must integrate over the source.   This could in principle be done
directly in the source plane, but would prove difficult because 
the magnification diverges as the source position approaches the caustic.  
We therefore employ the method suggested by Bennett \& Rhie (1996), of integrating
in the image plane, where the magnification is well behaved.  In this case, the 
magnification is simply given by,
$$
A = {{\sum_{i=1}^{n} {\Omega}_i}\over{\Omega}_s},
\eqn\fmag
$$
where ${\Omega}_i$ is the area of image $i$, and ${\Omega}_s = \pi\rho^2$ 
is the area of the source.  The difficulty, therefore, lies in finding the
images, which in general are scattered throughout the image plane.  
Fortunately, for $b \le 0.3$, and $|\zeta| \le 0.06$, the
images are all confined to a thin annulus of radius $|\zeta| = 1$ for both
the binary and single lens.  Using this method, we now construct contours
of $\epsil$ as a function of position in the source plane, $\zeta$, for
a range of binary separations.  In Figure {\four} we show contours of 
$\epsil =  0.03$ and $0.10$ for $b= 0.10, 0.15, 0.20, 0.25$.  Since
the binary detection probability depends more strongly on $b$ than
$q$, we have included only the results for $q=1.0$.  The results for other
mass ratios are qualitatively similar.  

As in $\S 3$, we calculate the binary detection probability by integrating over
lens trajectories in the intervals $0 \le \theta \le 2\pi$ and $0 \le \beta \le 1$.   In
Figure {\five} we show the binary detection probability as a function of $b$ for 
the range $0.10 \le b \le 0.25$.  Figure {\five} shows that finite source effects
decrease the detection probability for $b < 0.25$, and render the binary 
virtually undetectable for $b \lsim 0.1$, confirming the analytic estimate 
below equation {\caus}.

\chapter{Extreme Microlensing and Binary Detection}

There exists a small subclass of events in which it is possible to measure both 
the proper motion, $\mu$, and the projected Einstein radius, $\tilde r_e$, from
ground-based measurement alone, 
and thus determine the mass of the lens.  These extreme microlensing
events (EMEs) have been discussed by Gould (1996b), and are characterized by a 
very high maximum
magnification $A_{\rm{max}}$.  The basic requirement to be able to measure $\mu$ and
$\tilde r_e$ is, $A_{\rm{max}} \gsim 200$, and,
$$
\beta \lsim \rho.
\eqn\require
$$
We now determine whether it is possible to 
detect the presence of a binary of a given separation  for these types of events.  
Gould (1996b) found that the
typical source stars for EMEs are solar-type stars.  For a source of physical radius 
$R \sim R_{\odot}$,  the dimensionless radius is 
$\rho \simeq 0.001$ for $M=0.4 M_{\odot}$, $\dol = 4\, \kpc$, and
$\dos = 8\, \kpc$.  From equation \caus, we estimate that the binary will still
be detectable as long as $w \ge 0.001$.  Using the relation \caussize, it is apparent
that binaries of separation $b \gsim 0.05$ will be detectable in all EMEs.  In fact,
binaries of somewhat smaller separations will still be detectable, since
as mentioned in \S 6, the requirement that $w \ge \rho$ is only approximate.  Furthermore,
to measure $r_e$, the sampling rate for EMEs must 
be very high, typically one observation per minute,
with photometric precisions of $\le 1\%$.  With such observations,
one would be able to detect binaries of much smaller separations, perhaps down to $b \sim .01$.

{\bf Acknowledgements}:
We would like to thank P. Sackett for several stimulating discussions.
This work was supported in part by grant AST 94-20746 from the NSF.  

\endpage
\Ref\Alard{Alard, C. 1996, in Proc. IAU Symp. 173 (Eds. C.S. Kochanek, J.N. Hewitt), 
in press (Kluwer Academic Publishers)}
\Ref\Alcock{Alcock, C., et al.\ 1995, ApJ, 454, L125}
\Ref\Alcock{Alcock, C., et al.\ 1996, ApJ, submitted (astro-ph=9606012)}
\Ref\Aubourg{Aubourg, E., et al.\ 1995, A\&A, 301, 1}
\Ref\bandr{Bennett, D.\ \& Rhie, S.\ 1996, preprint (astro-ph=9603158)}
\Ref\bandg{Boutreux, T.\ \& Gould, A.\ 1996, ApJ, 462, 705}
\Ref\kamion{Buchalter, A.\ \& Kamionkowski, M.\ 1996, ApJ, submitted (astro-ph=9604144)}
\Ref\fandm{Fischer, D.\ \& Marcy, G.\ 1992, ApJ, 396, 178}
\Ref\me{Gaudi, B. S., \& Gould, A. 1996, ApJ, submitted (astro-ph=9601030)}    
\Ref\gone{Gould, A.\ 1992, ApJ, 392, 442}
\Ref\gtwo{Gould, A.\ 1994, ApJ, 421, L71}
\Ref\gtwo{Gould, A.\ 1995a, ApJ, 441, L21}
\Ref\gtwo{Gould, A.\ 1995b, ApJ, 447, 491}
\Ref\gtwo{Gould, A.\ 1996a, PASP, 108, 465} 
\Ref\gthre{Gould, A.\ 1996b, ApJ, submitted (astro-ph=9603148)}
\Ref\gandw{Gould, A. \& Welch, D.\ 1996, ApJ, 464, 212}
\Ref\ktg{Kroupa, P., Tout, C., \& Gilmore, G.\ 1991, 251, 293}
\Ref\mandb{Marcy, G.\ \& Benitz, K.\ 1989, ApJ, 344, 441}
\Ref\nem{Nemiroff, R.\ J.\ \& Wickramasinghe, W.\ A.\ D.\ T.\ 1994, ApJ, 424, L21}
\Ref\refsdal{Refsdal, S.\ 1966, MNRAS, 134, 315}
\Ref\reid{Reid, N.\ 1991, AJ, 102, 1428}
\Ref\ogle{Udalski, A., et al. 1994, ApJ, 436, L103}
\Ref\wandm{Witt, H., \& Mao, S. 1994, ApJ, 430, 505}
\Ref\witt{Witt, H.\ 1990, A\&A, 236, 311}
\refout
\endpage
%\figout
\endpage
\endpage
\endpage
\bye